\documentclass[amsmath,amssymb,aps,twocolumn]{revtex4}

\newcommand{\wt}{\widetilde}
\newcommand{\lp}{\left}
\newcommand{\rp}{\right}
\newcommand{\EE}{{\cal E}}
\newcommand{\OO}{{\cal O}}

\begin{document}

\begin{flushright}
CERN-PH-TH/2008-142
\end{flushright}
\vskip -0.9cm


\title{Non Pauli-Fierz Massive Gravitons}

\author{Gia~Dvali$^{a,b}$, Oriol~Pujol\`as$^b$ and Michele~Redi$^c$}

\address{$^a$CERN, Theory Division, CH-1211 Geneva 23, Switzerland}
\address{$^b$Center for Cosmology and Particle Physics,
New York University, New York, NY 10003 USA}
\address{$^c$Institut de Th\'eorie des Ph\'enom\`enes Physiques, EPFL, CH-1015, Lausanne, Switzerland}

\begin{abstract}
We study general Lorentz invariant theories of massive gravitons.
We show that, contrary to the standard lore, there exist
consistent theories where the graviton mass term violates
Pauli-Fierz structure. For theories where the graviton is a
resonance this does not imply the existence of a scalar ghost if
the deviation from Pauli-Fierz becomes sufficiently small at high
energies. These types  of mass terms are required by any
consistent realization of the DGP model in higher dimension.
\end{abstract}

\maketitle

{\em Introduction:}
The current accelerated expansion of the
Universe is arguably the most relevant observation in modern
cosmology. The fact that it might be signaling a failure of
General Relativity (GR) at large distances is a compelling idea
that motivates the investigation of large distance modifications
of gravity. Since GR is the unique consistent theory of massless
spin 2 field, whose low energy limit is fixed by the principle of
invariance under general coordinate transformations, any infrared
modification of GR requires some sort of mass for the graviton.

The subject of massive spin-2 fields traces back to the classical work of
Pauli-Fierz \cite{pf}. At quadratic order around a flat background there
are two possible mass terms compatible with  Lorentz invariance,
\begin{equation}\label{NPF}
V=\frac {M_4^2}4 \left[m_{PF}^2  h^{\mu\nu} \lp(  h_{\mu\nu} -  \eta_{\mu\nu} h \rp) %
+ m^2   h^2\right]~,
\end{equation}
where $h_{\mu\nu}$ is the metric fluctuation and $h=h_\mu^\mu$.
The celebrated Pauli-Fierz (PF) mass term \cite{pf} corresponds to
$m=0$. Due to the mass $h_{\mu\nu}$ propagates more degrees of
freedom than in the massless case (2). In the PF case the graviton has five
polarizations while when also $m\ne0$ there is an extra scalar.
This can be seen as follows. The scalar longitudinal degree of
freedom of the graviton can be isolated performing the diffeomorphism \cite{nima},
\begin{equation}
h_{\mu\nu}= \widehat{h}_{\mu\nu}+ \frac
{2\,\partial_\mu\partial_\nu \phi}{m_{PF}^2}~. \label{diffeo}
\end{equation}
The mass term \eqref{NPF} is not invariant and generates a kinetic
term for the scalar,
\begin{equation}
- \delta{\cal L}=M_4^2  \Big[\frac {m^2}{m_{PF}^2}\, (\Box
\phi)^2+\phi (\partial_\mu\partial_\nu \widehat h^{\mu\nu}-\Box
\widehat h)
 + \dots\Big].\label{higherderivative}
\end{equation}
For $m=0$ the scalar obtains a (healthy) kinetic term by mixing
with the transverse polarizations of the graviton \cite{nima}. For
$m\ne 0$, $\phi$ acquires a four derivatives kinetic term
signaling the presence of a second scalar. The higher
derivative structure in fact implies that this extra degree of freedom
is a ghost. For this reason graviton mass with structure different
from PF are normally discarded.

The purpose of this Letter will be to show that this conclusion does not hold
in general and that non-PF massive gravity theories can be consistent if the graviton is a resonance. This corresponds to the case where the mass parameters depend on the energy scale which is natural
from the point of view of extra dimensions. In fact the only Lorentz invariant non-linear theories
of massive gravity  known to date resort to extra dimensions.
The prototype of these types of theories is provided by the Dvali-Gabadadze-Porrati (DGP)
model \cite{dgp} where five dimensional gravity is localized on a codimension one defect by means
of a brane kinetic term for the graviton. An important feature of these geometrical models is that the
tensor structure depends on the number of extra dimensions. For instance one can show that the
DGP model describes a graviton resonance with a mass term of the PF form \cite{pf}.
The higher dimensional generalizations of DGP \cite{gregorygia} require a different mass term,
and essentially for this reason they were thought to be inconsistent. Recently however
a counterexample was presented where a ghost free model in 6 dimensions was
constructed \cite{us}. As we will discuss this can be understood as the first
realization of a consistent non-PF massive gravity theory.\\[-2mm]

{\em Consistent Massive Gravitons:} The basic reason why it might
be possible to find a ghost free theory of non-PF massive
gravitons is due to the fact that such a theory is equivalent to a
scalar-tensor theory. From the point of view of the irreducible
representation of the Lorentz group the spectrum can be decomposed
into scalars and PF massive gravitons. Since they belong to
different representations it is consistent with Lorentz symmetry
to change the UV behavior in Eq. (\ref{higherderivative}) that is
responsible for the presence of ghosts. To show how this works,
let us consider mass terms of the form \eqref{NPF} where we assume
in general that $m_{PF}$ and $m$ are scale-dependent functions.
Adopting a parameterization similar to \cite{gia06}, by Lorentz
invariance this implies that $m_{PF}$ and $m$ must be functions of
the d'Alembertian, $\Box$.  The connection with scalar-tensor
theories  can be realized integrating in a scalar field in the
action,\\[-4mm]
\begin{eqnarray}\label{scalarTensor}
{\cal L}&=& {M_4^2 \over4}\lp[  h^{\mu\nu}(\EE  h)_{\mu\nu} -
m_{PF}^2 h^{\mu\nu}  \lp(  h_{\mu\nu} -  h \eta_{\mu\nu}
\rp)\rp]\cr &+& \alpha\, \phi h+\beta\, \phi^2 + h^{\mu\nu}
 \, T_{\mu\nu}
\end{eqnarray}
(here, $(\EE h)_{\mu\nu}=\Box h_{\mu\nu}+\dots$ denotes the
linearized Einstein tensor). This is equivalent to the non-PF
massive graviton (\ref{NPF}) if $\alpha^2 = m^2 M_4^2 \, \beta$.
%
We now perform the diffeomorphism (\ref{diffeo}) followed by a rescaling of the metric,
\begin{equation}
h_{\mu\nu} = \wt h_{\mu\nu} + \phi  \eta_{\mu\nu} +
{2\, \partial_\mu\partial_\nu\over m_{PF}^2} \phi ~.
\label{shift}
\end{equation}
where the shift of the metric is chosen in order to remove the
kinetic mixing term $\phi \, (\EE h)$ in Eq.
(\ref{higherderivative}). One obtains,
\begin{eqnarray}\label{scalarTensor2}
{\cal L}&=& {M_4^2 \over4}\lp[\wt h^{\mu\nu}(\EE \wt h)_{\mu\nu}
- \wt h^{\mu\nu} m_{PF}^2 \lp(\wt  h_{\mu\nu} - \wt h
\eta_{\mu\nu} \rp)\rp]\cr %
&+& \alpha\, \phi \left(\wt h + 4 \phi + \frac {2\,\Box
\phi}{m_{PF}^2}\right) + \beta\, \phi^2 + {3 M_4^2\over 2} \,
m_{PF}^2 \, \phi \, \wt h \cr &+& \big(\wt h_{\mu\nu} + \phi
\eta_{\mu\nu} +
{2\, \partial_\mu\partial_\nu\over m_{PF}^2} \phi  \,\big) T^{\mu\nu} ~. %
\end{eqnarray}
By choosing $\alpha=-\frac 3 2 M_4^2 m_{PF}^2$,
we cancel the mixing between $\phi$ and $h$ and we arrive at,
\begin{eqnarray}\label{scalarTensor}
{\cal L}&=& {M_4^2 \over4}\lp[ \wt h^{\mu\nu}(\EE \wt h)_{\mu\nu}
-  \wt h^{\mu\nu} m_{PF}^2 \lp(  \wt h_{\mu\nu} -  \wt h
\eta_{\mu\nu} \rp)\rp]\cr %
&+& {M_4^2 \over 2} \,\phi {\cal O} \phi
+ \left( \wt h_{\mu\nu} + \phi\, \eta_{\mu\nu}+ \frac {2
\,\partial_\mu\partial_\nu}{m_{PF}^2}\phi \right) \,  T^{\mu\nu}
~,\\
&&{\rm with}\quad{\cal O}\equiv{9\over2}\,\frac {m_{PF}^4}{m^2} -
3 (\Box+2 m_{PF}^2) \,. \label{O}
\end{eqnarray}
Therefore the non-PF graviton can be rewritten in terms of a PF
graviton and a scalar decoupled from each other, in agreement with
\cite{gia06}. Note that an analogous decomposition is impossible
for the scalar longitudinal component of a PF graviton because
this polarization belongs to the same representation of the
Lorentz group.  The propagator of non-PF graviton can be
reconstructed using Eq. (\ref{shift}) in terms of PF and scalar
propagators.
We can also read off from \eqref{scalarTensor} the amplitude for
the exchange between two conserved sources,
\begin{equation}
{\cal A}=\frac 1 {M_4^2} \left(\frac {T_{\mu\nu}T'^{\mu\nu}-(1/3)
TT'} {\Box - m_{PF}^2} + {1\over2}\frac {T T'}{{\cal O}}\right)
\label{scalaramplitude}
\end{equation}
where the first contribution corresponds to the massive PF
graviton contribution and the second to the scalar exchange.

More in general, for massive gravitons gauge invariance does not
demand conservation of $T_{\mu\nu}$.  However, the non-conserved
part is still constrained and has to vanish  in the limit of zero
graviton mass.
At the quantum level this brings the following subtlety. For a
non-conserved source the coupling of the form ${\phi
\partial_{\mu} \partial_{\nu} \over \bar m^2} T^{\mu\nu}$ can generate
a higher derivative kinetic term for  $\phi$,
\begin{equation}
\label{correlator} {\partial_{\mu} \partial_{\nu} \phi \over \bar
m^2} \langle T^{\mu\nu} T^{\alpha\beta} \rangle {\partial_{\alpha}
\partial_{\beta} \phi \over \bar m^2}~.
\end{equation}
For the conserved  $T_{\mu\nu}$, the correlator  $\langle
T^{\mu\nu} T^{\alpha\beta} \rangle $ is proportional to the
transverse projector and the higher derivative kinetic term
(\ref{correlator}) vanishes. For non-conserved sources, the
structure can be non-zero but, as we mentioned, the  non-conserved
part of the source must go to zero at least as $\bar m^2$ itself.
This means that the higher derivative kinetic term will be
suppressed by the cutoff of the theory and the resulting ghost
pole is automatically at the cutoff. Notice that the same
correlator will generate exactly the same type of higher
derivative mass term for the helicity zero polarization of the PF
graviton, with $\bar m^2 = m^2_{PF}$. So the absence of any ghost
pole below the cutoff requires that the divergence of $T_{\mu\nu}$
is controlled by $m^2_{PF}$.  Since our scalar couples to
$T_{\mu\nu}$ with  $\bar m^2 =m^2_{PF}$, the absence of ghosts at
the quantum level in the PF graviton automatically implies the
absence of similar ghosts in $\phi$.

Coming back to Eq. (\ref{O}), in any theory with \emph{constant}
$m_{PF}$ and $m$ the scalar propagator has a pole
at,\\[-5mm]
\begin{equation}
p^2=-\frac {3 m_{PF}^4} {2 m^2}+ 2 m_{PF}^2~.\\[-2mm]
\end{equation}
From the sign of the kinetic term in Eq.  (\ref{O}) this pole is
always a ghost, with positive or negative mass-squared depending
on the sign of $m^2$. One interesting feature is that no matter
how small the deviation from PF structure is, the amplitude
\eqref{scalaramplitude} has the same tensor structure of massless
gravity in the UV. This is because the ghost couples to
$T_\mu^\mu$ with the same strength as the longitudinal
polarizations of the massive graviton and therefore it exactly
cancels its contribution at high energy. For $m\to0$ the ghost
becomes heavy and decouples from the theory.

When $m$ and $m_{PF}$ are scale dependent and different however
then one does not necessarily have ghosts in the spectrum due to
the positive contributions in Eq. \eqref{O}. Let us now discuss
how the consistency of the theory actually constrains the masses.
In general, absence of ghosts demands the spectral decomposition
of the scalar and spin-2 amplitudes in Eq. (\ref{scalaramplitude})
to be positive definite. A necessary condition is that the
coefficient of the scalar contribution is strictly positive since
a negative value (corresponding to repulsion) could only be
provided by ghosts. This strongly restricts the form that $m_{PF}$
and $m$ can take.

We can obtain different constraints considering the infrared (IR) and ultraviolet
(UV) behavior of the amplitude. In the UV the positivity of the scalar amplitude requires,\\[-5mm]
\begin{equation}\label{condition}
m^2(\Box) < {3\over2} { m_{PF}^4(\Box) \over \Box }
\end{equation}
in the limit $\Box\to \infty$. If this condition is violated then a ghost
appears. Note that in this case ${\cal O}$ scales as $\Box$ in the UV so this ghost
corresponds a 4D scalar bound state. The condition above can also be understood
as the fact that non-PF term does not generate ghosts as long as it is sufficiently
subleading at high energies. This also agrees with the effective field theory intuition that
small deviations from PF are acceptable \cite{nima}.

In the IR we can parameterize the behavior of the tensor and the scalar as,\\[-5mm]
\begin{equation}\label{parameterizing}
m_{PF}^2\approx A_2 \, (-\Box)^{a_2}~,  %
\qquad\; %
-\OO \approx  A_0\, (-\Box)^{a_0}~.
\end{equation}
Since we are interested in large distance modifications of gravity
we consider $a_{0,2}<1$ so that these terms become dominant with
respect to the kinetic term at large distances. As discussed in
\cite{gia06}, the spectral decomposition requires that the scalar
and spin 2 amplitude should not vanish at zero momenta. This
implies $a_{0,2}\ge 0$ and $A_{0,2}>0$. Note that if the
positivity constraint is violated in the IR the amplitude has a
branch cut corresponding to a continuum of ghosts. From this and
\eqref{O}, one finds that the healthy forms of $m^2$
in the IR are
\begin{equation}\label{ir}
  m^2(\Box)=\frac 3 4 \frac {A_2(-\Box)^{a_2}} {1 -
  A_0/(6A_2)(-\Box)^{a_0-a_2}}~.
\end{equation}
This automatically classifies all possible cases in 3 families:
(i) $a_0>a_2$, corresponding to the scalar being `lighter' than
the tensor ($-\OO \ll m_{PF}^2$) in the IR. These cases have
$m^2/m_{PF}^2
\to 3/4 $.
(ii) $a_0 = a_2$, which includes the geometrical models discussed
below. In this cases ${m^2/ m_{PF}^2}$ approaches a constant with
a value outside the range between $0$ and $3/4$.
(iii) $a_0 < a_2$. In this case, $m^2\to (9/2) (m_{PF}^4/\OO)$,
so one has $-m^2 \ll m_{PF}^2 \ll -\OO$.\\[-1mm]

{\em Geometrical Realizations:}
We now turn to the geometrical realization of gravitons with non-PF structure.
As mentioned earlier a natural arena for theories of massive gravitons is higher
dimensional theories of gravity. Following \cite{dgp,gregorygia}, the addition
of a kinetic term for the graviton on the lower dimensional defect insures the
existence of a 4D regime,
\begin{equation}\label{4+n}
S=\frac {M_4^2} 2 \int R_4+ \frac {M_{4+n}^{2+n}}2 \int R_{4+n}~.
\end{equation}

The connection with massive gravity theories can be made manifest
computing the boundary effective action obtained by integrating
out the bulk degrees of freedom. Expanding the action \eqref{4+n}
around flat space, one can choose a gauge
where the brane is located at $\vec y =0$ in the $n-$dimensional
transverse space. The induced metric fluctuation perceived by
brane observers then reduces to the 4-dimensional components
evaluated on the brane, $h_{\mu\nu}(\vec y=0)$. One can further
fix the gauge so that the graviton propagator takes the form\\[-4mm]
\begin{eqnarray}
G_{MNPQ}&=&\frac 1 {M_{4+n}^{2+n}}\frac 1 {p^2+ {\vec q}\;{}^2} \times\\
\nonumber &\times&\big[\; \frac 1 2 \eta_{MP}\eta_{NQ}+\frac 1 2
\eta_{M Q} \eta_{NP}-\frac 1 {2+n} \eta_{MN}\eta_{PQ}\big]
\end{eqnarray}
where $M,N\dots$ denote $4+n$ dimensional indices and $p$ (${\vec
q}$) the 4- ($n$-) dimensional components of the momentum.
Neglecting the 4D term in \eqref{4+n}, the amplitude between two
brane localized sources is,
\begin{equation}
\int d^4p\;G(p)\Big(\wt T_{\mu\nu}(p) \wt T'^{\mu\nu}(-p)-\frac
{\wt T(p) \wt T'(-p)} {2+n} \Big) \label{A}
\end{equation}
where $\wt T_{\mu\nu}(p)$ is the Fourier transform of the source
and
\begin{equation}
G(p)=\frac 1 {M_{4+n}^{2+n}}\int \frac {d^n q}{p^2+{\vec
q}\;{}^2}~.
\label{propscalar}
\end{equation}
This integral is divergent for $n\ge 2$ and therefore requires
some regularization. Introducing a momentum cutoff
$\Lambda$,\footnote{The expansion \eqref{propexp} holds for $n$
even (for $n=2$, the first term is absent). For $n$ odd the
non-local term is $p^{n-2}$.}
\begin{equation}
G(p)\simeq \frac
{\Omega_{n}}{M_{4+n}^{2+n}}\Big[\,{\Lambda^{n-2}\over n-2}+ ...+
(-1)^{n\over2}p^{n-2}\log \frac p {\Lambda}+... \Big],
\label{propexp}
\end{equation}
where $\Omega_{n}=2\pi^{n/2}/\Gamma(n/2)$ and local terms of the
form $(p/\Lambda)^{2k}\Lambda^{n-2}$ with $k=1,2...$ that are
present for $n>4$ have been omitted. The displayed
cutoff-independent non-local term generates the higher dimensional
Newtonian potential $\sim 1/r^{n+1}$.

The amplitude \eqref{A} can be derived from the boundary effective
action,\\[-5mm]
\begin{equation}
\int d^4x h^{\mu\nu} G^{-1}(\Box) \left[h_{\mu\nu}-\frac 1
{n-2}\eta_{\mu\nu} h \right]+h_{\mu\nu} T^{\mu\nu} \label{Seff}
\end{equation}
where $h_{\mu\nu}$ is the metric measured by the brane observer.
The addition of the brane localized kinetic term $\int R_4 $ then
leads to a massive gravity  action where the bulk provides a scale
dependent mass term for the graviton. For the case $n=1$, the 5D
DGP model,  this corresponds to a resonance with Pauli-Fierz mass.
For $n>1$ the tensor structure of the higher dimensional theory is
encoded  in a non-PF mass term.

From \eqref{Seff} and \eqref{propexp} we can identify,\\[-5mm]
\begin{equation}\label{mPF(p)}
m_{PF}^2= \frac {1}{ M_4^2\,G(\Box)} , \qquad %
m^2=\frac {n-1}{n-2} \,m_{PF}^2  ~. \\[-2mm]
\end{equation}
The necessity of the non-PF mass for $n>1$ can also be understood
from the Kaluza-Klein decomposition of these theories. For $n>1$,
aside from the tower of gravitons there is a tower of spin-0 states,
which are encoded in the scalar $\phi$ that we integrated in.

We are now in the position to see why the $n>1$ theories propagate
ghosts, as first shown in \cite{dubov,gregshif}. This is just a
consequence of $m$ not obeying \eqref{condition} in the UV. More
precisely, from \eqref{scalaramplitude} one derives that there is
light ghost pole with a mass\\[-6mm]
\begin{equation}
m_{ghost}^2\approx - \frac1 {2\Omega_n} \, \frac {n+2}{n-1}
\,\frac{n-2}{ \Lambda^{n-2}} \, \frac {M_{4+n}^{2+n}}{M_4^2} ~.
\end{equation}
The most important feature of this formula is the dependence on
the inverse cutoff (for codimension 2 this becomes logarithmic).
While a heavy ghost can be consistent within an effective field
theory approach this formula shows that the ghost is light. For
this reason, higher dimensional generalizations of DGP were
believed to be inconsistent.

Following our general analysis this conclusion is a manifestation
of the fact that $m_{PF}$ and $m$ have the same momentum
dependence so that $m$ never becomes subleading. As we have seen
this is not mandatory because the extra scalars can, consistently
with 4D Lorentz invariance, couple differently than the massive
spin 2 states. In order to possibly avoid the ghost $m$ and
$m_{PF}$ should scale differently in the UV. This was explicitly
realized in the ``cascading DGP model'' \cite{us}. In the 6
dimensional case considered in that paper a 3-brane is embedded
within a codimension 1 brane, each with their own induced gravity
term\\[-5mm]
\begin{equation}
S=\frac {M_4^2} 2  \int R_4+ \frac {M_5^3} 2 \int R_5 +\frac
{M_6^4} 2 \int R_{6}  \nonumber \\[-2mm]
\end{equation}
In a certain limit of the model, with the addition of a tension
$\lambda$ on the 3-brane, it was found that the 4D boundary
effective action reduces to,\\[-5mm]
\begin{eqnarray}
\label{boundaryEffAct2} L_4&=  &  - {M_5^3\over2}\;
h^{\mu\nu}\sqrt{-\Box}\lp(  h_{\mu\nu} -
      h \eta_{\mu\nu} \rp) %
     - 3 M_5^3\, \pi \sqrt{-\Box} \,   h \cr
  &  +&{M_4^2\over4} \;  h^{\mu\nu}(\EE  h)_{\mu\nu}
  + {9 \lambda \over 4m_6^2 } \pi \Box \pi %
 + h^{\mu\nu} \, T_{\mu\nu} ~,
\end{eqnarray}
where $m_6=M_6^4/M_5^3$. The scalar $\pi$ corresponds to the brane
bending mode of the 4-brane whose $4D$ kinetic term arises from
the tension. To see the relationship with our general analysis one
can integrate out $\pi$ using its equation of motion, $\pi=-(2
M_5^3 m_6^2)/(3 \lambda)\;(-\Box)^{-1/2} h$. Substituting into
the action generates a non-PF mass,\\[-5mm]
\begin{equation}\label{masses}
m_{PF}^2 =\frac { 2 M_5^3}{M_4^2} \sqrt{-\Box} \qquad , \qquad m^2
=-4{M_5^6 m_6^2\over M_4^2\, \lambda}  ~. \\[-2mm]
\end{equation}
Since  $m_{PF}$ grows linearly with energy $m$ becomes subleading
in the UV  and the condition \eqref{condition} is satisfied for $3
\lambda> 2 M_4^2 m_6^2$. This reproduces the bound of Ref.
\cite{us}.\\[-3mm]

{\em Outlook:} Before concluding we wish to speculate on other
realizations of graviton resonances with non-PF structure.  One
possible direction is the generalization of the cascading DGP
model to higher codimensions. To realize this we need to consider
a tower of DGP kinetic terms embedded into each other,\\[-5mm]
\begin{equation}
S=\frac {M_4^2} 2 \int R_4+ \frac {M_5^3} 2 \int R_5 +\dots +\frac
{M_{4+n}^{2+n}}2 \int R_{4+n} \\[-2mm]
\end{equation}
At very large distances physics is dominated by the
$4+n-$dimensional kinetic term and asymptotically can be described
by a non-PF resonance with parameters given by Eq. (\ref{mPF(p)}).
As in the codimension 2 case the lower dimensional kinetic terms
generate brane localized ghosts which can be studied using the
method of the boundary effective action. This can be derived
integrating out the bulk degrees of freedom step by step starting
from the highest codimension. To obtain a consistent theory one
should insure that at each step one obtains a consistent massive
gravity theory on the lower dimensional defect. This should be
achieved by introducing sources, such as the tension in
codimension 2, which render the non-PF mass term appropriately
subleading in the UV.

A different possibility is to consider theories with a non-integer
number of extra-dimensions, which corresponds to a fractal extra
space.
Formally for a brane observer these theories can be obtained by
analytic continuation of eqs. (\ref{A}), (\ref{propscalar}).
In the range $0<n<2$, \eqref{propscalar} is finite and one obtains
$m_{PF}^2=p_c^n\, p^{2-n}$ with \\[-6mm]
%
%
\begin{equation}
p_c^n=\frac {\Gamma(n/2) \sin(n\,\pi/2) }{ \pi^{1+\frac n 2} } ~
\, \frac {M_{4+n}^{n+2}} {M_4^2} \label{fractional} \nonumber \\[-4mm]
\end{equation}
and $m^2=(n-1)/(n-2)\;m_{PF}^2$.
This reproduces the momentum dependence of the theories considered
in \cite{gia06} but in order to interpret these as boundary
effective actions arising from a geometry we also continue the
tensor structure. For the case $n<1$ the scalar amplitude
corresponds to the propagation of a continuum of light ghosts.
This can be readily seen because the scalar exchange  in Eq.
(\ref{scalaramplitude}) is negative.  This rules out the range
$n<1$. For $1<n<2$, the masses above satisfy the constraints in
the IR \eqref{ir} but not in the UV (\ref{condition}). The ghost
pole is at $p^2=p_c^2 [(1+n/2)/(n-1)]^{1/n}$ so that for $n$ close
to 1 it is heavy and hence could be accepted within an effective
field theory approach.
From a geometrical point of view these theories can be realized
through a bulk space which is fractal. At quadratic order they can
be defined using a lattice as in \cite{Hill} (see however \cite{nima2} for
difficulties at interacting level). It is appealing that this
construction only generates $n>1$. Theories with $1<n<<2$ are also
the most interesting from a phenomenological point of view as they
could be tested by future lunar ranging experiments \cite{gia06}.
We leave the detailed construction of these generalizations of the
DGP model to future work.

\emph{Acknowledgements} We would like to thank D. Blas and G.
Gabadadze for useful discussions. This work has been supported in
part by the NSF grant PHY-0245068, by the David and Lucile Packard
Foundation Fellowship for Science and Engineering and by DURSI
under grant 2005 BP-A 10131.

\end{document}